\documentclass[rnote,bibyear]{aa}
\usepackage[varg]{txfonts}
\usepackage{graphicx}
\usepackage{hyperref}

\begin{document}

   \title{Ghosts of Milky Way's past: the globular cluster ESO\,37-1 
          (E\,3)
          \thanks{This research note is based on observations made 
                  with the ESO VLT at the Paranal Observatory, under 
                  the program 078.D-0186 and includes data gathered 
                  with the 6.5 metre Magellan Telescopes located at 
                  Las Campanas Observatory, Chile (program ID 
                  CHILE-2015A-029).}
          \thanks{Appendix A is available in electronic form at 
                  http://www.aanda.org. Tables of the individual 
                  photometric measurements are available at CDS via 
                  anonymous ftp to cdsarc.u-strasbg.fr (130.79.128.5) 
                  or via
                  http://cdsweb.u-strasbg.fr/cgi-bin/qcat?J/A+A/XX/XX}
         }
   \author{R. de la Fuente Marcos\inst{\ref{inst1}}
           \and C. de la Fuente Marcos\inst{\ref{inst1}}
           \and C. Moni Bidin\inst{\ref{inst2}}
           \and S.~Ortolani\inst{\ref{inst3}}
           \and G. Carraro\inst{\ref{inst4}, \ref{inst3}}
          }
   \authorrunning{de la Fuente Marcos et al.}
   \titlerunning{The globular cluster ESO\,37-1 (E\,3)
                }
   \offprints{R. de la Fuente Marcos, \email{rauldelafuentemarcos@gmail.com}
             }
   \institute{Apartado de Correos 3413, E-28080 Madrid, Spain\label{inst1}
              \and
              Instituto de Astronom\'{\i}a, Universidad Cat\'olica del Norte,
              Av. Angamos 0610, Antofagasta, Chile\label{inst2}
              \and
              Dipartimento di Fisica e Astronomia, Universit\`a degli Studi di Padova,
              Vicolo dell'Osservatorio 3, I-35122, Padova, Italy\label{inst3}
              \and
              European Southern Observatory, Alonso de Cordova 3107,
              Casilla 19001, Santiago 19, Chile\label{inst4}
             }
   \date{Received 22 May 2015 / Accepted 5 July 2015}

   \abstract
      {In the Milky Way, most globular clusters are highly conspicuous objects 
       that were found centuries ago. However, a few dozen of them are faint, 
       sparsely populated systems that were identified largely during the second 
       half of the past century. One of the faintest is ESO\,37-1~(E\,3) and as 
       such it remains poorly studied, with no spectroscopic observations 
       published so far although it was discovered in 1976.
       }
      {We investigate the globular cluster E\,3 in an attempt to better 
       constrain its fundamental parameters. Spectroscopy of stars in the field 
       of E\,3 is shown here for the first time.  
       }
      {Deep, precise $VI$ CCD photometry of E\,3 down to $V$$\sim$26~mag is 
       presented and analysed. Low-resolution, medium signal-to-noise ratio 
       spectra of nine candidate members are studied to derive radial velocity 
       and metallicity. Proper motions from the UCAC4 catalogue are used to 
       explore the kinematics of the bright members of E\,3.
       }
      {Isochrone fitting indicates that E\,3 is probably very old, with an age 
       of about 13~Gyr; its distance from the Sun is nearly 10~kpc. It is also 
       somewhat metal rich with [Fe/H]=$-0.7$. Regarding its kinematics, our
       tentative estimate for the proper motions is ($\mu_{\alpha}~\cos\delta,
       \mu_{\delta})=(-7.0\pm$0.8, 3.5$\pm$0.3)~mas~yr$^{-1}$ (or a tangential 
       velocity of 382$\pm$79~km~s$^{-1}$) and for the radial velocity  
       45$\pm$5~km~s$^{-1}$ in the solar rest frame.
       }
      {E\,3 is one of the most intriguing globular clusters in the Galaxy. 
       Having an old age and being metal rich is clearly a peculiar combination, 
       only seen in a handful of objects like the far more conspicuous NGC\,104 
       (47\,Tucanae). In addition, its low luminosity and sparse population make 
       it a unique template for the study of the final evolutionary phases in 
       the life of a star cluster. Unfortunately, E\,3 is among the most elusive 
       and challenging known globular clusters because field contamination 
       severely hampers spectroscopic studies.
       }


   \maketitle

  \section{Introduction}
     In the Milky Way, most globular clusters are highly conspicuous and rich stellar systems that were discovered centuries ago (see e.g. 
     Hanes 1980; Ashman \& Zepf 2008). However, a few dozen objects are faint and sparsely populated; they were identified predominantly 
     during the second half of the past century (see e.g. Harris 1976, 1996). Most globular clusters were born with the Galaxy, but others 
     were captured and today are found associated with more or less prominent stellar streams (see e.g. Lynden-Bell \& Lynden-Bell 1995; 
     Freeman \& Bland-Hawthorn 2002; Forbes \& Bridges 2010). 

     One of the faintest known globular clusters is ESO\,37-1 (E\,3), which  was discovered by A. Lauberts in Chamaeleon in 1976 (Lauberts 
     1976); it is still a poorly studied object and no spectroscopic observations have been published. Harris (2010) gives the position of 
     the cluster as $\alpha$(h:m:s, J2000) = 9:20:57.07, $\delta$(\degr:\arcmin:\arcsec, J2000) = $-$77:16:54.8 or $l$ = 292\fdg270, $b$ = 
     $-$19\fdg020, with $M_V$ = $-$4.12 mag, [Fe/H] = $-$0.83 dex and located at a heliocentric distance of 8.1 kpc or galactocentric 
     distance of 9.1 kpc. Sarajedini et al. (2007) point out that E\,3 appears to be $\sim$2 Gyr younger than NGC 104 (47 Tucanae), a very 
     conspicuous 12$\pm$1 Gyr old (McDonald et al. 2011) globular cluster.

     Here, we present and analyse deep, precise $VI$ CCD photometry of E\,3 down to $V$$\sim$26 mag. In addition, we study low-resolution, 
     medium signal-to-noise ratio spectra of nine candidate cluster members. This research note is organised as follows. Section 2 is 
     devoted to photometry. The spectroscopy is discussed in Section 3. The proper motions of possible bright members of the cluster are 
     investigated in Section 4. In Section 5, a polar path analysis is presented. Our results are discussed and our conclusions summarised 
     in Section 6.

  \section{Photometry}
\begin{figure}
        \centerline{\hbox{
                    \includegraphics[width=\columnwidth]{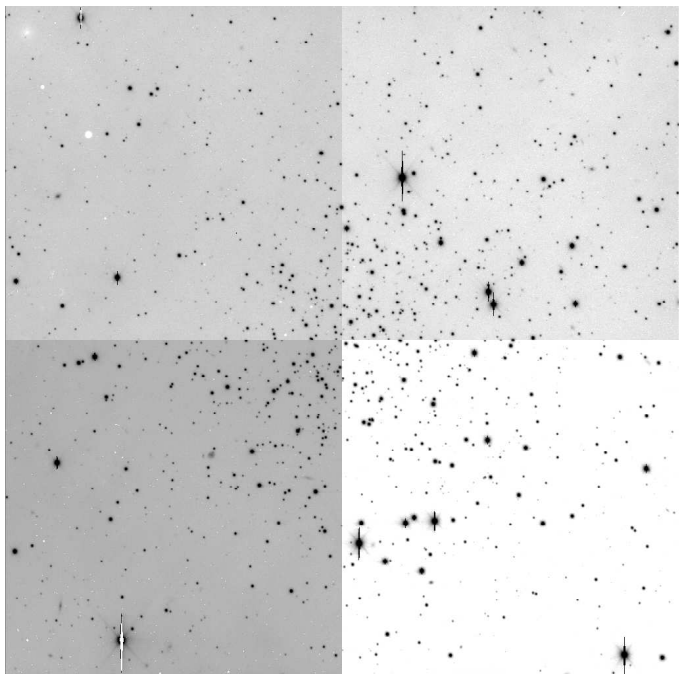}
                   }}
        \caption{$V$ 1000-second exposure in the field of E\,3. North is up, east to the left, and the field is 6\farcm8 on a side.
                }
        \label{example}
     \end{figure}

     On the night of 13 November 2006, E\,3 was observed in the $VI$ filter bands with the Very Large Telescope UT2 Kueyen and the FORS1 CCD
     camera. Four exposures were taken (10 s and 1000 s in both $V$ and $I$) on that night, which had nominal photometric conditions and an 
     average seeing of 0\farcs8, at an airmass in the range 1.77--1.82. The camera has a scale of 0\farcs2 per pixel and an array of 
     2048$\times$2048 pixels. An example of one of the 6.8$\times$6.8 arcmin$^2$ images of E\,3 is shown in Fig.~\ref{example}. An adjacent
     field was observed under the same conditions, offset by half a degree in the west direction. The standard IRAF\footnote{IRAF is 
     distributed by the National Optical Astronomy Observatory, which is operated by the Association of Universities for Research in 
     Astronomy (AURA) under a cooperative agreement with the National Science Foundation.} routines were applied to reduce the raw images. 
     Using the psf-fitting routines of DAOPHOT and ALLSTAR (Stetson 1994) in the IRAF environment, we measured instrumental magnitudes for 
     all the stars in the field. These magnitudes were converted to the standard system using the stars in common with Veronesi et al. 
     (1996) as local standards; transformations were subsequently utilised to tie both the cluster and the field photometry to them (tables 
     including the individual photometric measurements are available at CDS). 
%
%
%
%

\begin{figure}
       \centerline{\hbox{
                    \includegraphics[width=\columnwidth,angle=0]{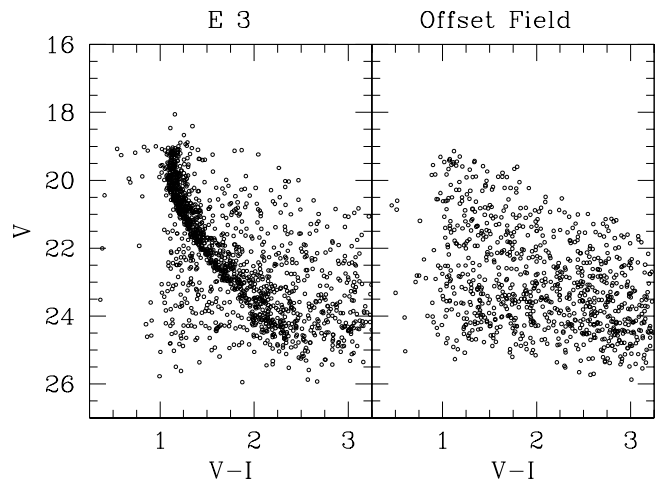}
                  }}
                \caption{CMD for E\,3 (left panel) and the comparison field (right panel).}
       \label{cmds}
    \end{figure}

     The resulting colour magnitude diagrams (CMD) are shown in Fig.~\ref{cmds}, for both the cluster (left panel) and the field (right 
     panel). One can readily see that we can cover the cluster main sequence (MS) from the turn-off point (TO) down to $V$$\sim$25 mag, two 
     magnitudes fainter than Veronesi et al. (1996). The photometry reaches approximately the same depth as the archival HST images of E\,3 
     discussed in Lan et al. (2010). For the purpose of this study, we combined our photometry with that in Gratton \& Ortolani (1987) to 
     have a complete magnitude coverage, and to also sample the evolved part of the CMD (see below). Photometric errors for stars with 
     $V$$\approx$20 mag are $\sim$0.03~mag (0.05 in $V-I$); for $V$$\approx$23 mag they amount to $\sim$0.10~mag (0.15 in $V-I$).
%
%
%
%

     In order to estimate star membership in the cluster, we made use of both the photometry of the cluster and the accompanying offset 
     field. First of all, we performed a completeness analysis, both for the cluster and for the field, which is summarised in 
     Table~\ref{complete}. This analysis was carried out in the standard way (see Baume et al. 2007) by adding a number of stars per 
     different magnitude bin and randomly across the field, and then computing the ratio of recovered to injected stars, where the recovered 
     stars result from the reduction of the added-star-image under the same conditions as in the original image. Looking at 
     Table~\ref{complete}, one can conclude that our photometry is complete down to $V$$\sim$25 mag ($I$$\sim$24 mag) and that the cluster 
     is affected by some crowding.
%
%
     \begin{table}
      \centering
       \caption{Results of the completeness test}
       \begin{tabular}{ccc}
          \hline\hline
                       & Cluster & Field \\
          \hline
           $V$         & comp    & comp  \\
           (mag)       & (\%)    & (\%)  \\
          \hline
           20.0$-$23.0 &  100    & 100   \\
           23.5$-$24.5 &   90    & 100   \\
           24.5$-$25.0 &   64    &  90   \\
           25.0$-$25.5 &   17    &  55   \\
           $\geq$ 25.5 & $\geq$0 &  20   \\
          \hline
           $I$         & comp    & comp  \\
           (mag)       & (\%)    & (\%)  \\
          \hline
           20.0$-$22.7 &  100    & 100   \\
           22.7$-$24.0 &   72    & 100   \\
           23.0$-$24.5 &   24    &  68   \\
           24.5$-$25.0 &   10    &  32   \\
       \end{tabular}
       \label{complete}
     \end{table}
%
%

     The completeness results have been subsequently applied to perform a statistical cleaning of the cluster CMD making use of the field, 
     as described in Carraro \& Costa (2007), with the aim of isolating statistically the genuine cluster population. Summarising, the 
     statistical technique we employ works as follows. We pick  an offset field star and look into the cluster CMD for the closest star in 
     colour and magnitude using a search ellipse, and we remove this star from the cluster star list. This procedure is repeated for all the 
     stars in the offset field. We refer the reader to the cited paper for the full details on this procedure. In the following analysis we 
     make use of this resulting {\it \emph{clean}} CMD.
%
%
    \begin{figure}
       \centerline{\hbox{
                    \includegraphics[width=\columnwidth]{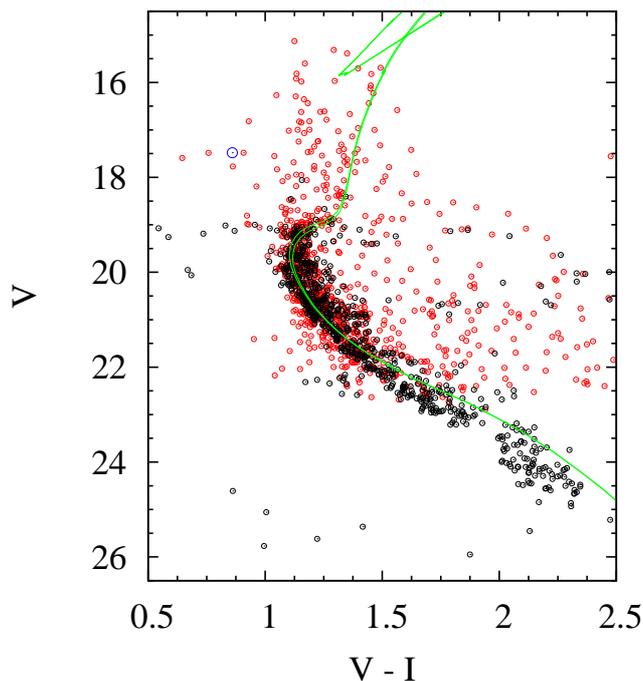}
                  }}
                \caption{CMD for the clean sample (black circles). Photometry from Gratton \& Ortolani (1987) is also shown (red circles). 
                         Two isochrones (12--13 Gyr, $Z$ = 0.003) from the PARSEC evolutionary tracks (Bressan et al. 2012) are shown. The 
                         blue straggler V1, an SX Phoenicis variable star (Mochejska et al. 2000) and a probable cluster member, is shown as 
                         a blue circle.
                         }
       \label{clean}
    \end{figure}
%
%

     Figure \ref{clean} shows the CMD for the clean sample as black circles, and data from Gratton \& Ortolani (1987) as red circles. Two 
     isochrones with $Z$ = 0.003, and 12 and 13 Gyr of age are used to perform a main-sequence fitting, giving ($V-M_V$)$_0$ = 15.07 mag and 
     $E(V-I)$ = 0.47; the values of $A_V$ and $A_I$ from Schlafly \& Finkbeiner (2011) are 0.93 mag and 0.51 mag, respectively. Using these 
     values, our estimate for the cluster heliocentric distance is 10.3$\pm$0.9 kpc. As the studied field is restricted to 6\farcm8 on a 
     side, we disregard estimating the cluster absolute magnitude and structural parameters from the photometry presented here.

  \section{Spectroscopy}
\begin{figure}
        \centerline{\hbox{\includegraphics[width=\columnwidth]{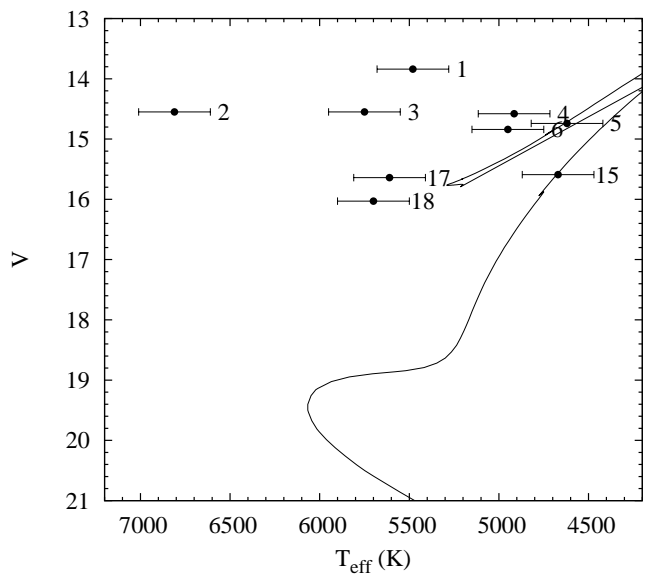}
                   }}
                   \caption{Comparison between the position of the stars studied spectroscopically (see Table~\ref{t_spec}) in the 
                            temperature-magnitude diagram with an assumed cluster sequence, $Z$ = 0.003 and age of 13 Gyr, from the PARSEC 
                            evolutionary tracks (Bressan et al. 2012).
                            }
                   \label{f_spec}
     \end{figure}

     Intermediate-resolution ($R$$\approx$7000) spectra of nine stars in the field of E\,3 were collected at the 6.5m Baade Telescope, Las 
     Campanas Observatory, on 28--29 January and 22 February 2015. The IMACS spectrograph was used at f/4, with the 1200+26.7 grating and 
     $0\farcs75$-wide slit. The resulting spectral range 5200--7000~{\AA} was covered by four chips, with one short gap between each pair of
     chips. Exposure times varied between 240s and 900s, depending on target magnitude. The four spectral sections were reduced individually 
     with standard IRAF routines, extracted and normalised. Before merging them, the offsets between the four independent wavelength 
     calibration solutions were measured on a twilight solar spectrum and corrected. We measured the radial velocity (RV) of the target 
     stars cross-correlating (see Tonry \& Davis 1979) their spectra with synthetic templates drawn from the library of Coelho et~al. 
     (2005). The results were corrected to heliocentric velocities and for zero-point offsets, estimated from the position of the strong 
     telluric band observed at 6850--7000~{\AA}. The errors were computed as the quadratic sum of the most relevant sources of uncertainty, 
     namely the cross-correlation, the wavelength calibration, and the RV zero-point correction (for further details, see Carraro et al. 
     2007). We estimated the temperature of the targets fitting the wings of the H$_\alpha$ line with the same synthetic spectra (see 
     Fuhrmann et~al. 1994) following the procedure applied in Moni Bidin et~al. (2010). We varied the gravity and metallicity of the 
     template models and we thus verified that, as expected, the results were largely insensitive to these parameters. The uncertainty 
     associated with the adopted $\chi^2$ minimisation is likely an underestimate of the true errors because it takes into account only 
     random errors (spectral noise), while sources of systematics such as continuum definition and selection of data points for the fit must 
     also play a role. From the comparison of estimates obtained with different model parameters, normalisation, and fit range we estimated 
     that the true uncertainties should likely be of the order of 150--200~K. The results are given in Table~\ref{t_spec}, while in 
     Fig.~\ref{f_spec} we compare the position of the stars in the temperature-magnitude diagram with the cluster sequence. Four stars, 
     indicated with ``field (hot)'' in the last column of the table, are too hot for an old stellar cluster and they will not be considered 
     further. Unfortunately, we found no spectral feature suitable for a reliable gravity estimate at our low resolution. The degeneracy 
     between gravity and metallicity in the fit of the spectra could therefore not be fully broken and we derived the best-fit metallicity 
     when $\log{g}$ was varied in the range 0--4~dex. The metallicity of star 4 resulted [Fe/H]$>-$0.35 for any value of $\log{g}$, and the 
     best-fit solution of star 5 indicated that it is most likely a dwarf star ($\log{g}>4$). Their cluster membership could thus be 
     excluded and they are flagged with ``field (met.)'' in Table~\ref{t_spec}. The metallicity of star 17 was also too high 
     ([Fe/H]$\approx$0.0) to be of interest here. Object 6 is likely a red giant ($\log{g}\approx1.8$) with [Fe/H]=$-0.65\pm0.15$, while the 
     spectrum of target 15 is compatible with either a metal-poor giant ($\log{g}\approx2$, [Fe/H]$=-0.7\pm0.2$), or a metal-rich dwarf.
%
%
     \begin{table}
        \tabcolsep 0.10truecm
        \begin{center}
           \caption{Spectroscopic results.}
           \label{t_spec}
           \begin{tabular}{c| c c r c c l}
              \hline\hline
               ID & RA(J2000)  & Dec(J2000)              & RV            & T$_\mathrm{eff}$ & $V$   & Notes        \\
                  & (h:m:s)    & (\degr:\arcmin:\arcsec) & (km~s$^{-1}$) & (K)              & (mag) &              \\
              \hline
                1 & 09:20:48.0 & $-$77:15:15             & 6$\pm$9       & 5500             & 13.84 & field (hot)  \\
                2 & 09:20:55.9 & $-$77:19:01             & $-$23$\pm$3   & 6800             & 14.55 & field (hot)  \\
                3 & 09:20:41.9 & $-$77:18:47             & $-$35$\pm$3   & 5750             & 14.55 & field (hot)  \\
                4 & 09:20:32.1 & $-$77:16:25             & $-$21$\pm$4   & 4900             & 14.58 & field (met.) \\
                5 & 09:20:47.3 & $-$77:18:48             & 6$\pm$4       & 4600             & 14.74 & field (met.) \\
                6 & 09:20:31.2 & $-$77:16:33             & 45$\pm$3      & 4950             & 14.84 & Member       \\
               15 & 09:19:23.5 & $-$77:14:39             & 44$\pm$5      & 4650             & 15.59 & Member       \\
               17 & 09:19:41.6 & $-$77:08:07             & 41$\pm$4      & 5600             & 15.64 & field        \\
               18 & 09:19:17.2 & $-$77:09:40             & 28$\pm$6      & 5700             & 16.03 & field (hot)  \\
              \hline
           \end{tabular}
        \end{center}
     \end{table}
%
%
%
%
%
%

     In conclusion, only two stars in the observed sample (6 and 15) have both temperature and metallicity fully compatible with cluster 
     membership, and their RVs differ by only 1~km~s$^{-1}$. Target 17 also has similar RV, but -- although its higher temperature is still 
     compatible with a horizontal branch cluster star -- its high metallicity points to a field star. Therefore, our estimated value for 
     the radial velocity of E\,3 is 45$\pm$5~km~s$^{-1}$ in the solar rest frame.

  \section{Proper motions}
     The two most robust candidates for membership in E\,3 have proper motions in UCAC4 (Zacharias et al. 2013). Star 6 is UCAC4\,064-008502 
     with ($\mu_{\alpha}~\cos\delta, \mu_{\delta})$ = ($-6.4\pm$1.9, 3.3$\pm$1.7)~mas~yr$^{-1}$. Star 15 (UCAC4\,064-008471) has consistent 
     proper motions of ($-7.6\pm$2.0, 3.7$\pm$2.3)~mas~yr$^{-1}$. Star 17 (UCAC4\,065-008924) has somewhat inconsistent proper motions, 
     ($-1.6\pm$3.5, 1.1$\pm$5.4)~mas~yr$^{-1}$. Assuming that stars 6 and 15 are part of the cluster, our tentative estimate for the proper 
     motions is ($\mu_{\alpha}~\cos\delta, \mu_{\delta})$ = ($-7.0\pm$0.8, 3.5$\pm$0.3)~mas~yr$^{-1}$. A tangential velocity of 
     382$\pm$79~km~s$^{-1}$ is derived from the inferred cluster proper motion.
    
     The globular cluster E\,3 is a low-luminosity stellar system with very few bright stars in its central regions. The closest star to the 
     assumed centre of the cluster (VizieR, J2000 09:20:57.1 h:m:s, $-$77:16:55 \degr:\arcmin:\arcsec) with proper motions in UCAC4 is 
     40{\arcsec} from the centre. The closest putative member could be UCAC4\,064-008511 with ($-7.0\pm$4.3, 4.1$\pm$13.9)~mas~yr$^{-1}$, 
     63{\arcsec} from the centre. Stars 6 and 15 are located 1\farcm4722 and 5\farcm6358, respectively, from the centre. In contrast, 
     star 17 is located 9\farcm0846 from the cluster centre. Van den Bergh et al. (1980) estimated the core and tidal radii of E\,3 at 
     4.4~pc (1\farcm51) and 24.7~pc (8\farcm49), respectively. Being extra-tidal is another argument against star 17 having membership in 
     E\,3.
%
%
     \begin{figure}
        \centerline{\hbox{\includegraphics[width=\columnwidth]{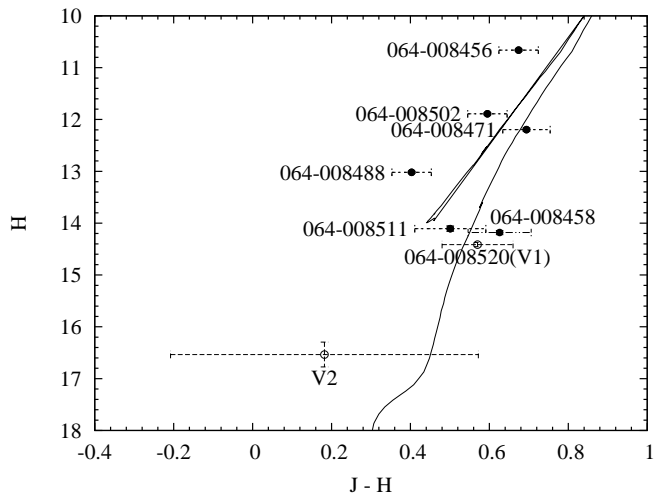}
                   }}
                   \caption{2MASS near-infrared CMD of stars within 8\farcm5 from the cluster centre and with proper motions within
                            2$\sigma$ of our estimate for the cluster (black circles, labels are the UCAC4 designations). Variable stars V1 
                            and V2 from Mochejska et al. (2000) are also indicated (empty circles).
                            }
                   \label{f_pm}
     \end{figure}
%
%

     Figure \ref{f_pm} shows the 2MASS near-infrared CMD of selected stars. The objects plotted have proper motions compatible with those of
     stars 6 (064-008502) and 15 (064-008471) and are located within the cluster tidal radius. The sources V1 and V2 are putative cluster 
     members from Mochejska et al. (2000), but V2 has no proper motions in UCAC4 and V1 has rather incompatible proper motions, 
     (0.7$\pm$5.7, 22.2$\pm$5.7) mas~yr$^{-1}$.

  \section{Polar paths}
     Most globular clusters were born with the Galaxy, but some others were accreted. Each Galactic globular cluster and satellite dwarf 
     galaxy orbits the Galaxy in a plane including the object's present position and the Galactic centre. This reference plane can be 
     defined as the instantaneous or osculating orbital plane of the object. Following Lynden-Bell \& Lynden-Bell (1995) we consider the 
     geometric loci of all possible normals to the galactocentric radius-vector for a particular object (see Appendix A for mathematical 
     details). This defines the set of possible orbital poles for that object, globular cluster, or dwarf galaxy.       

     For E\,3, the usual heliocentric galactic coordinates $(l, b) = (292\fdg270, -19\fdg020)$ translate into galactocentric galactic 
     coordinates $(l_{g}, b_{g}) = (232\fdg550, -16\fdg474)$. We  solved Eq. \ref{poles} using a Monte Carlo approach and evaluated the 
     statistical significance of all the crossing points of polar path intersections with that of E\,3 using the Kulldorff statistics
     (Kulldorff 1997). All studies coincide in classifying E\,3 as a metal-rich globular cluster. For this globular cluster subset the most 
     relevant intersection (180$\sigma$) appears to be located at $X$ = $-$0.42, $Y$ = 0.50 or ($l, b$) = (310$^{\circ}$, +35$^{\circ}$), 
     see Fig. \ref{polarpaths}. In these diagrams a separation of 0.1 is equivalent to $\sim6^{\circ}$. The poles in Fig. \ref{polarpaths} 
     are all within 2$^{\circ}$ of the average pole. Prospective members of this dynamical family of objects include, besides E\,3, the 
     second most massive Galactic globular cluster NGC~104 (47 Tucanae), NGC 6171 (M 107), and NGC 6362. Additional members may be Terzan 2 
     and NGC 6652. All of them have ages in the range 12.8$\pm$0.6 Gyr and metallicity [Fe/H] in the range $(-0.69, -1.02)$. If we compare
     the polar paths of this group of objects with those of objects commonly associated with the Sagittarius stream (see Figs. 
     \ref{polarpaths} and \ref{polarpathsSAGDEG}), the dispersion of the poles is even smaller in this case.
%
%
     \onlfig{
     \begin{figure*}
        \centerline{\hbox{\includegraphics[width=\linewidth]{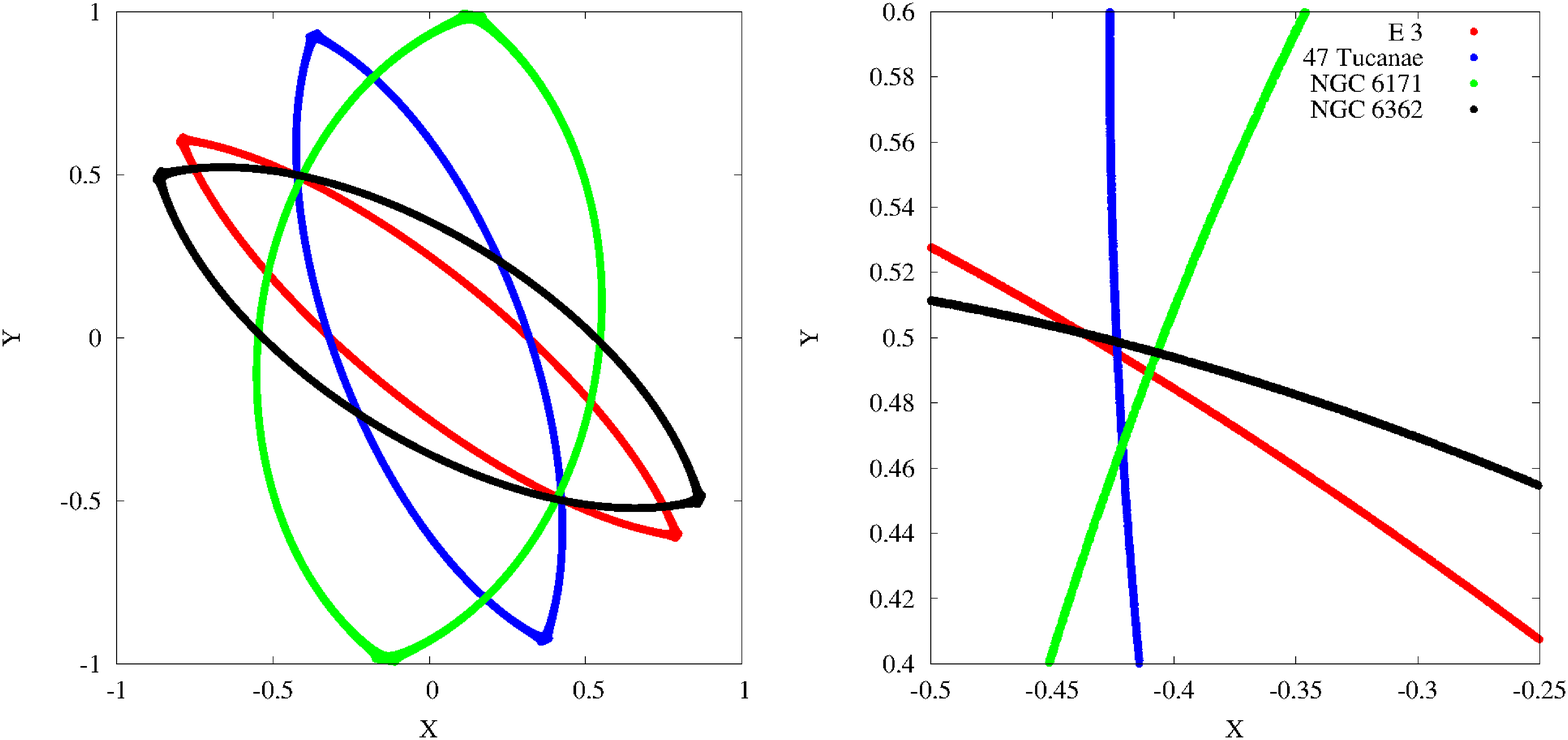}
        }}
        \caption{(Left panel) Polar paths of objects believed to be part of the E\,3 stream. (Right panel) Close-up.
             }
        \label{polarpaths}
     \end{figure*}
     }
%
%
%

  \section{Discussion and conclusions}
     Clearly, more observations are required to  investigate this interesting cluster further. In particular, detailed spectroscopy of the 
     brightest stars can provide important information on the actual metallicity of the cluster and its radial velocity. However, our 
     preliminary spectroscopic study has revealed that this cluster is profusely contaminated with field stars. On the other hand, its low 
     luminosity and sparse population make it a unique template for the study of the final evolutionary phases in the life of a star 
     cluster. An extensive spectroscopic study can be used to measure the radial velocity dispersion that in turn can be utilised to 
     constrain the dynamical state of this object. In a cluster with a high binary fraction like E\,3 (e.g. Veronesi et al. 1996), binary 
     orbital motions can mask small velocity dispersions. Multi-epoch observations can be used to separate single from binary systems. 
     Unfortunately, E\,3 is among the most elusive and challenging known globular clusters because field contamination severely hampers 
     spectroscopic studies.  

     The globular cluster E\,3 does not have a clear horizontal branch and that makes its distance and age determination rather uncertain. 
     Our values differ significantly from those in Sarajedini et al. (2007), namely 8.1 kpc and 10 Gyr, respectively. The value of $A_V$ 
     used in this study is nearly 17\% lower than those used in previous works. Our isochrone fitting indicates that E\,3 is probably very 
     old, with an age of about 13~Gyr and its heliocentric distance is nearly 10~kpc. We have confirmed that it is also somewhat metal rich 
     with [Fe/H]=$-$0.7. For its kinematics, our tentative estimate for the proper motion is ($\mu_{\alpha}~\cos\delta, \mu_{\delta})$ = 
     ($-7.0\pm$0.8, 3.5$\pm$0.3)~mas~yr$^{-1}$. A tangential velocity of 382$\pm$79~km~s$^{-1}$ is derived from our determination of the 
     cluster proper motion and a radial velocity of 45$\pm$5~km~s$^{-1}$ is obtained from our spectroscopy, in the solar rest frame. The 
     E\,3 stellar system is one of the most intriguing globular clusters in the Galaxy. Its old age and high metallicity is a peculiar 
     combination, only seen in a handful of objects like the far more conspicuous NGC\,104 (47\,Tucanae) that may share a stellar stream 
     with it.  

  \begin{acknowledgements}
     We thank the anonymous referee for the constructive and useful report. CMB acknowledges support from FONDECYT through regular 
     project 1150060. In preparation of this research note, we made use of the NASA Astrophysics Data System and the ASTRO-PH e-print 
     server. This research has made use of the SIMBAD and VizieR databases operated at CDS, Strasbourg, France. This publication makes use 
     of data products from the Two Micron All Sky Survey, which is a joint project of the University of Massachusetts and the Infrared 
     Processing and Analysis Center/California Institute of Technology, funded by the National Aeronautics and Space Administration and the 
     National Science Foundation.
  \end{acknowledgements}

  \bibliographystyle{aa}

  \Online

  \begin{appendix}
     \section{Polar path technique analysis}
\begin{figure*}
        \centerline{\hbox{\includegraphics[width=\linewidth]{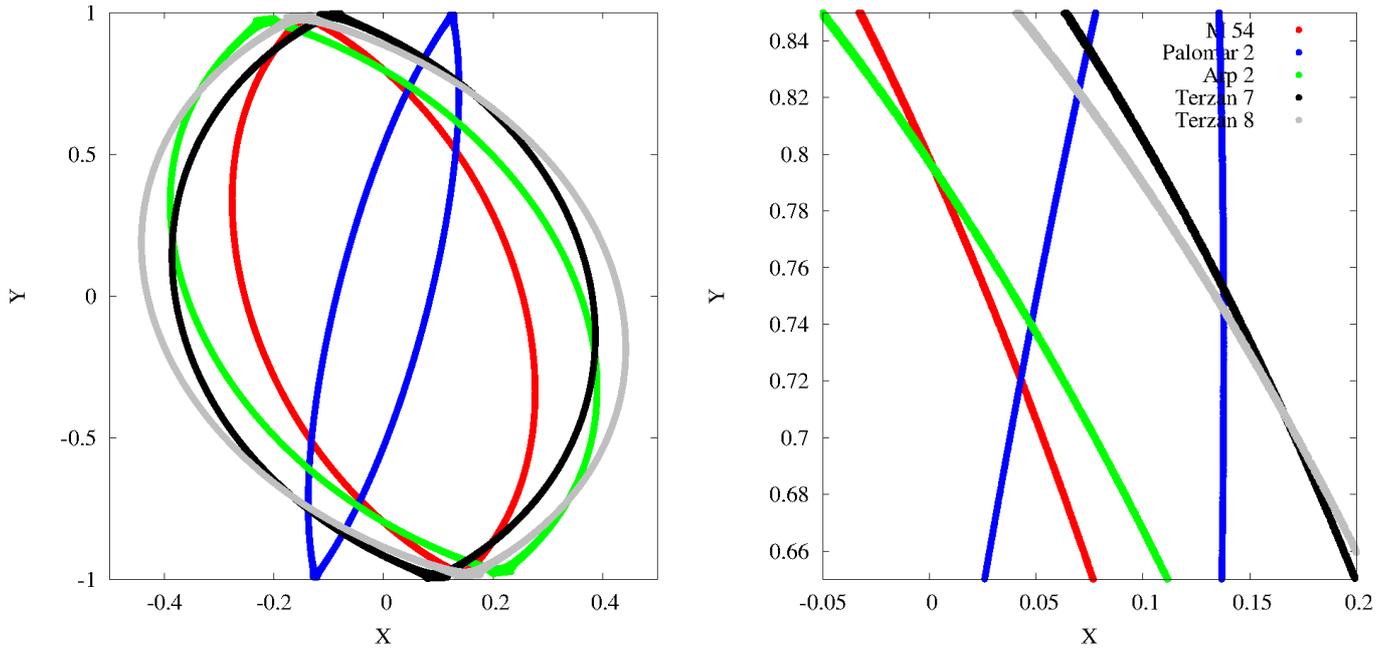}
        }}
        \caption{(Left panel) Polar paths of objects believed to be part of the Sagittarius stream, streams 8a and 8b in Table 2 of 
                 Lynden-Bell \& Lynden-Bell (1995). (Right panel) Close-up.
             }
        \label{polarpathsSAGDEG}
     \end{figure*}

        Each Galactic globular cluster and satellite dwarf galaxy orbits the Galaxy in a plane including the object's present position and 
        the Galactic centre. This reference plane can be defined as the instantaneous orbital plane of the object. Following Lynden-Bell \& 
        Lynden-Bell (1995) we consider the geometric loci of all possible normals to the galactocentric radius-vector for a particular 
        object. This defines the set of possible orbital poles for that object, globular cluster, or dwarf galaxy. We consider a system of 
        galactic coordinates on the unit sphere ($x^2 + y^2 + z^2 = 1$) centred on the Galactic centre; the coordinates of any object in 
        that system are given by
        \begin{equation}
           (x, y, z) = (\cos l_{g} \ \cos b_{g}, \sin l_{g} \ \cos b_{g}, \sin b_{g}) \,,
                         \label{gal}
        \end{equation}
        where $l_{g}$ and $b_{g}$ are the galactocentric (not the regular heliocentric) galactic longitude and latitude, respectively. The 
        galactocentric radius-vectors for an object and its pole are perpendicular, therefore
        \begin{equation}
           x \cos l_{g} \cos b_{g} + y \sin l_{g} \cos b_{g} + z \sin b_{g} = 0 \,.
              \label{dot}
        \end{equation}
        Finding $z(x, y)$ from Eq. \ref{dot}, replacing it on the expression of the unit sphere, and reorganising we obtain
        \begin{equation}
           x^2 \ \left(1 + \frac{\cos^2 l_{g}}{\tan^2 b_{g}}\right) + y^2 \ \left(1 + \frac{\sin^2 l_{g}}{\tan^2 b_{g}}\right) +
             2 x y \frac{\sin l_{g} \cos l_{g}}{\tan^2 b_{g}} = 1 \,. 
              \label{locus}
        \end{equation}
        If we now consider the Lambert azimuthal equal-area projection given by the expressions
        \begin{equation}
           (X, Y) = (x/\sqrt{1 + z}, y/\sqrt{1 + z}) \,, 
             \label{lambert}
        \end{equation}
        with $z = 1 - X^2 - Y^2$, Eq. \ref{locus} can be rewritten as
        \begin{equation}
           (2 - X^2 - Y^2) \left(X^2 + Y^2 + (X \cos l_{g} + Y \sin l_{g})^2 \frac{1}{\tan^2 b_{g}}\right) = 1 \,. 
            \label{poles}
        \end{equation}
        This expression provides the positions of all the possible poles associated with the object. The intersection (if any) of polar 
        paths gives the putative pole of a group of (perhaps) dynamically related objects. Once the pole has been found, its position on the 
        sky relative to the Galactic centre is given by $(l_P, b_P) = (\arctan(Y_P/X_P), \arcsin(1-X_{P}^{2}-Y_{P}^{2}))$, see Eqs. 
        \ref{gal} and \ref{lambert}. Polar paths of objects possibly associated with E\,3 are shown in Fig. \ref{polarpaths}. For 
        comparison, those of objects widely regarded as part of the Sagittarius stream, streams 8a and 8b in Table 2 of Lynden-Bell \& 
        Lynden-Bell (1995), are plotted in Fig. \ref{polarpathsSAGDEG}.
     \newpage
%
%
%
%
  \end{appendix}
\end{document}